\documentclass{icrc}

\usepackage{times}
\usepackage{graphicx} 

\begin{document}

\title{Distributed Data Acquisition System for Pachmarhi Array of
\v Cerenkov Telescopes}
\author[1]{P. N. Bhat}
\author[1]{B. S. Acharya}
\author[1]{V. R. Chitnis}
\author[1]{A. I. D'Souza}
\author[1]{P. J. Francis}
\author[1]{K. S. Gothe}
\author[1]{S. R. Joshi}
\author[1]{P. Majumdar}
\author[1]{B. K. Nagesh}
\author[1]{M. S. Pose}
\author[1]{P. N. Purohit}
\author[1]{M. A. Rahman}
\author[1]{K. K. Rao}
\author[1]{S. K. Rao}
\author[1]{S. K. Sharma}
\author[1]{B. B. Singh}
\author[1]{A. J. Stanislaus}
\author[1]{P. V. Sudershanan}
\author[1]{S. S. Upadhyaya}
\author[1]{B. L. V. Murthy}
\author[1]{P. R. Vishwanath}
\affil[1]{Tata Institute of Fundamental Research, Homi Bhabha Road,
Mumbai 400 005, India}

\correspondence{P. N. Bhat (pnbhat@tifr.res.in)}

\runninghead{Bhat {\it et al.}: Data acquisition system for PACT}
\firstpage{1}
\pubyear{2001}


\maketitle

\begin{abstract}
Pachmarhi Array of \v Cerenkov Telescopes (PACT) consists of
a 5$\times$5 array of \v Cerenkov telescopes deployed over an
area of 100 $m$ $\times$ 80 $m$, in the form of a rectangular matrix.
The experiment is based on atmospheric \v Cerenkov technique
using wavefront sampling technique. Each telescope consists of
7 parabolic mirrors mounted para-axially on an equatorial mount.
At the focus of each mirror a fast phototube is mounted. In this
experiment a large number of parameters have to be measured and
recorded from each of the 175 phototubes in the shortest possible
time. Further, the counting rates from each phototube as well as
the analog sum of the 7 phototubes from each telescope (royal
sum) need to be monitored at regular intervals during the run.
In view of the complexity of the system, the entire array is
divided into four smaller sectors each of which is handled by an
independent field signal processing centre (FSPC) housed in a
control room that collects, processes and records information from
nearby six telescopes that belong to that sector. The distributed
data acquisition system (DDAS) developed for the purpose consists
of stand-alone sector data acquisition system (SDAS) in each of the
four FSPC's and a master data acquisition system (MDAS). MDAS
running in the master signal processing centre (MSPC) records data
from each of the 25 telescopes. The data acquisition and monitoring
PCs (SDAS and MDAS) are networked through LAN. The entire real time
software for DDAS is developed in C under $linux$ environment. Most
of the hardware in DDAS are designed and fabricated in-house. The
design features and the performance of the entire system along with
some other auxiliary systems to facilitate the entire observations 
will be presented.
\end{abstract}

\section{Introduction}

Pachmarhi Array of \v Cerenkov Telescopes is a ground based atmospheric
\v Cerenkov experiment designed to study VHE $\gamma-$ray emission from
astronomical sources. This experiment is based on wavefront sampling 
technique and consists of an array of 25 telescopes spread over an area
of 80 $m$ $\times$ 100 $m$, at Pachmarhi in Central India. Spacing between
neighbouring telescopes is 20 $m$ in E-W direction and 25 $m$ in N-S 
direction. Each telescope consists of 7 para-axially mounted parabolic 
mirrors of diameter 0.9 $m$ with a fast phototube
at the focus of each mirror. Each telescope is equatorially mounted
and its movement is remotely controlled and monitored in the control room.
Some of the design details of this experiment are given by Chitnis et
al., 2001. Using the array, arrival time of \v Cerenkov shower front
and \v Cerenkov photon density are measured at various locations in the
\v Cerenkov pool generated by air showers. In this paper we discuss 
the design of data acquisition system used in this experiment.

\section{Experimental setup}

Pachmarhi array is divided into four subgroups or sectors of six telescopes 
each (see Figure 1). Each sector can be operated as an independent unit.
At the centre of each sector there is a station housing 
field signal processing centre (FSPC). Pulses from phototubes are brought to
the respective centre through low attenuation RG213 cables of length
$\sim$ 40 $m$ each. Since the array is split into the sectors, length of pulse
cable is reduced thereby minimizing the distortion and attenuation of weak 
pulses from phototubes during their transmission through cables. In these
type of experiments, it is necessary to preserve the shape and size of
the \v Cerenkov pulse as much as possible to improve the angular as well
as energy resolution of the experiment.
Pulses from individual mirrors are processed and the informations 
regarding pulse height and arrival time of shower front at mirrors is
recorded by FSPC. At the centre of the array there is a control room which 
houses master signal processing centre (MSPC). Information relevant to
entire array such as arrival time of shower front at individual telescopes,
absolute arrival time of the event $etc$ are recorded by the MSPC. 

 
Figure 2 shows the experimental setup schematically. High voltages of
individual phototubes are controlled through Computerized Automated
Rate Adjustment and Monitoring System (CARAMS). Orientation of telescope
mounts is controlled by Automated Computerized Telescope Orientation
System (ACTOS). Pulses from individual phototubes are brought to the
respective FSPC which processes the pulses and generates trigger. On
generation of trigger, information is recorded by $Linux$ PC. Once a trigger
is generated by any of the stations, MSPC in control room records
information relevant to entire array. In the following, we discuss details
of these subsystems.


\section{Auxiliary control systems}

Auxiliary control systems include systems for phototube high voltage control,
orientation of telescopes and a remotely controlled phototube shutter mechanism
called APES. Details of these subsystems are given below.

\subsection{Computerized Automated Rate Adjustment and Monitoring System 
(CARAMS)}

It is necessary to ensure that the gains of all the phototubes are more or less
equal by equalizing their counting rates due to NSB.
This is achieved by setting high voltages of individual phototubes
such that their count rates are approximately equal. Given the differences
in characteristics of individual phototubes as well as differences in 
reflectivities of mirrors, high voltage requirements differ from one
phototube to another. In addition, the counting  rates are sensitive 
to sky brightness, sky clarity, ambient temperature, ambient light
pollution $etc$, and hence there is a need
to adjust them every night. This process is quite time consuming. Hence
CARAMS is designed for adjusting individual rates and monitoring them.
This system consists of the microprocessor based 64 channel high voltage
divider unit (C.A.E.N. model SY170) which is controlled by CAMAC based
controller module (C.A.E.N. model CY117B). High voltage is given to the 
crate and output voltage from each channel is adjusted using a resistance
divider network and a micro-motor. CARAMS is developed for changing the 
phototube voltages as well as for reading them back using suitable CAMAC 
commands to the controller module. Rates from phototubes are measured 
using scaler modules and are available to CARAMS. 
Counting rates are measured at a few fiducial values encompassing the
required counting rate. A quadratic fit is used to define the variation
of counting rate as a function of applied high voltage for each phototube.
Then the required high voltage is computed using interpolation. The
required counting rate is then obtained by a few iterations around the
interpolated value. This procedure is fast and efficient. Using CARAMS, 
counting rates of all 175 phototubes can be adjusted in about 20 minutes.

\subsection{Automated Computerized Telescope Orientation System (ACTOS)}
  
During observations, it is necessary to ensure that all telescopes are
tracking the putative source accurately. Also the provision for fast
movement of telescopes while switching over from source to background 
observations or from one source to another is needed. Low cost Automated 
Computerized Telescope Orientation System (ACTOS) is designed to achieve 
this. All the telescopes are equatorially mounted and each telescope is 
independently steerable in both E-W and N-S direction within $\pm$45$^\circ$.
The hardware of ACTOS designed in-house for remotely controlling movement 
of these telescopes consists of a semi-intelligent closed loop feedback
system with built-in safety features. The heart of the system is an angular
position sensor which is basically a gravity based transducer called 
clinometer (see Figure 3). The clinometers whose DC outputs are linearly 
proportional to their offset angles from vertical (60 $mV/deg$) are used as 
absolute angle encoders. Two clinometers are used, one to get
telescope angle in N-S direction and other for E-W direction. These
clinometers are calibrated by aligning telescopes to stars at various
angles and measuring clinometer output voltages. Clinometer outputs   
are given to a low pass filter and an integration type ADC which is readout
by a host-PC. The motor controller, an interface between the host-PC and
the stepper motor, carries out the task of movement of stepper motors as
instructed by the host-PC. Variable slew speeds are used to decelerate 
the speeding telescope in steps. At present
three different speeds are used, $viz.$, fast (70 Hz), slow (30 Hz) and
tracking (7.561 Hz). The system can orient to the putative source  from an
arbitrary initial position with an 
accuracy of $\sim$ (0.003 $\pm$ 0.2). The source pointing is monitored at 
an accuracy of $\sim$0.05$^\circ$ and corrected in real time. For details
on ACTOS refer to Gothe {\it et al.} (2000)


\section{Distributed Data Acquisition System (DDAS)}

DDAS consists of a Sector Data Acquisition System (SDAS) in four stations
and a Master Data Acquisition System (MDAS) in the main control room.


\subsection{Sector data acquisition system (SDAS)}

SDAS is designed to process the pulses from phototubes, generate a trigger
and record the required information locally. Two types of data are recorded 
by SDAS. On event trigger, event arrival time, arrival time of the \v 
Cerenkov shower front and the photon density  at individual mirrors are 
recorded. On monitor trigger, which is variable but periodic, count rates 
from various mirrors as well as telescope trigger rates are recorded.
Block diagram of SDAS is shown in Figure 4. Pulses from phototubes are brought 
to the station through a low attenuation coaxial cables and given to linear 
fan-out unit which produces three replica of the input pulse. 
One set outputs from this module are given to Fan-in-fan-out unit 
which adds all the 7 analog pulses from phototubes of a 
telescope. These analog sums are called Royal Sums. Other set of outputs from 
fan-out unit is discriminated. One set of discriminator outputs is given 
to digital delay circuit, output of which form the TDC STOP's. The other set 
of discriminator outputs goes to CAMAC scalers which measure the count rates.
Third set of outputs from fan-out is given to integrating type ADCs. 
The royal sum pulses are discriminated at predetermined thresholds and event 
trigger is generated when there is any four-fold coincidence among six royal 
sums. Once an event trigger is generated, CAMAC controller initiates the data 
recording process. TDC start pulse is generated and individual TDC channels 
are stopped by appropriately delayed signals from delay circuit as 
mentioned before. So TDC 
readings are the measure of relative delay in arrival of shower front at 
various mirrors. Following a trigger ADC gate is generated and pulses from 
various mirrors are digitized by the ADCs. Finally the CAMAC controller 
interrupts $linux$ based PC which initiates the data recording process.
Recorded data are the $latch$ information which tells us which telescopes 
have participated in the trigger, absolute time of the event arrival correct 
to $\mu s$ in addition to TDC and ADC's. At present, ADC's and TDC's
corresponding to six peripheral mirrors of each telescope are stored. 

In addition to the above mentioned event data, monitor data are also recorded 
periodically. Monitor interrupts are generated at MSPC and fed to each of
FSPC's which initiates recording of data comprising of mirror rates and
royal sums along with absolute time from RTC. Monitor interrupts are
variable but periodic.

\subsection{Master Data Acquisition System (MDAS)}

MDAS records data relevant to the entire array. There are two types of data
recorded corresponding to two types of triggers as in the case of SDAS.
Royal sums from all the stations are brought to the control room.
Whenever an event trigger is generated at any station, MDAS CAMAC controller
is interrupted and the TDC delays, event arrival time from RTC, latch 
indicating the sectors participating in trigger, no. of events recorded 
in various stations are read and stored. Following each periodic monitor
interrupt, royal sum rates are recorded. Using data recorded by MDAS, 
data from stations can be collated off-line.

\subsection{Software}

The software developed for PACT can be classified into four groups.

\begin{enumerate}
\item  System software consisting of 
\begin{itemize}
\item[a.] Data acquisition software which
initiates the data recording process following top priority event
interrupt. It also records RTC latched by event trigger and starts reading
the latched data from various CAMAC modules. The data read are stored in
hard disc using double buffer scheme in order to reduce the system dead
time. Checksum word is used at the end of each event data to improve
the data reliability. Data acquisition software is divided into several
sub-tasks as follows :

\begin{itemize}
\item [(i)] Device drive module (DDM)
\item [(ii)] Device drive control module (DDCM)
\item [(iii)] Data display module (DDSP) and
\item [(iv)] Data server module
\end{itemize}

\item[b.] Monitoring interrupt servicing routine which latches and 
reads all the scalers which count the various phototube and royal sum
counting rates during the time interval between two consecutive
interrupts. Monitoring data are stored in a different file for off-line
analysis.
\end{itemize}

\item  Utility software like 
\begin{itemize}
\item[a.] software to preset and synchronize the various RTC's, 
\item[b.] software to feed counting rate information to CARAMS through 
serial port. This is accomplished by a System Manager PC in MSPC which 
receives monitoring data from the SDAS's through LAN and routing them to 
the PC running CARAMS.
\item[c.] on-line quick-look analysis program to check the data fidelity 
and other checks.
\end{itemize}

\item  Online monitoring and display routines like 
\begin{itemize}
\item[a.] read and display count and chance coincidence rates for all 
the sectors, 
\item[b.] read and display royal sum rates online. 
\end{itemize}
These tasks exploit the $Linux$ networking capabilities since all the 
PC's under $Linux$ are networked.

\item  Several off-line data handling routines which read and check each and 
every aspect of the data and produce a test report for each run, produce
various types of distributions to ensure the general health of the data.

\end{enumerate}

Software under (1) and (2) are developed in $C$ while  (3) and (4) are 
designed under IDL. All the software except CARAMS and ACTOS work under
$Linux$ while the latter work under DOS.

\section{Status of PACT}

PACT is fully operational since December 2000. All the subsystems are
functioning satisfactorily. Several sources including Crab nebula, Mkn 421,
Mkn 501, Geminga and 1ES 1426+428 have been observed and some preliminary
results on some of these are presented by Bhat {\it et al.}, 2001 and 
Vishwanath {\it et al.}, 2001.

\end{document}